\documentclass[preprint,showpacs,preprintnumbers,amsmath,amssymb]{revtex4}

\usepackage{graphicx}
\usepackage{dcolumn}
\usepackage{bm}

\usepackage{color,ulem}

\begin{document}
\title{Shaping the fluorescent emission by lattice resonances in plasmonic crystals of nanoantennas}
\author{G. Vecchi}
\author{V. Giannini}
\author{J. G\'{o}mez Rivas}
\email{rivas@amolf.nl}
\address{Center for Nanophotonics, FOM Institute AMOLF, c/o Philips
Research Laboratories, High Tech Campus 4, 5656 AE Eindhoven, The
Netherlands.}

\date{\today}

\begin{abstract}
We demonstrate that the emission of light by fluorescent molecules
in the proximity of periodic arrays of nanoantennas or plasmonic
crystals can be strongly modified when the arrays are covered by a
dielectric film. The coupling between localized surface plasmon
resonances and photonic states leads to surface modes which increase
the density of optical states and improve light extraction. Excited
dye molecules preferentially decay radiatively into these modes,
exhibiting an enhanced and directional emission.
\end{abstract}

\maketitle

Nanoparticles made of noble metals are the object of a renewed
scientific interest because of their remarkable optical properties,
primarily related to the excitation of localized surface plasmon
resonances (LSPRs)~\cite{Maier04, Lal07}. Advances in
nanofabrication techniques allow to shape nanoparticles into the
so-called optical antennas or nanoantennas. These techniques make
possible to tune the LSPRs over a wide spectral range, providing
control over the associated local field enhancement and far-field
scattering~\cite{Muhlschlegel05}. Several studies have focused on
the radiative properties of emitters in proximity to optical
antennas~\cite{Lakowicz05, Sandoghdar06, Mertens06, Krenn07, Tam07,
Muskens07, Bakker08}. The enhancement and quenching of the
fluorescence by emitters close to metal nanoparticles have been
described in terms of the balance between radiative and
non-radiative decay rates~\cite{Ford84, Feldmann05, Novotny06}.
These studies bring the twofold interest of, first, clarifying the
fundamental processes of light-matter interaction at a nanoscale
and, second, the possible implementation of nanoantennas to enhance
the efficiency of devices for lighting, photovoltaics, and sensing
applications~\cite{Polman08}. One limiting factor of metallic
structures in general are their inherent losses~\cite{Soref07}.
Therefore, the prediction that the electrodynamic interaction
between LSPRs of nanoparticles arranged in plasmonic crystals can
lead to mixed nanoparticle-lattice states or lattice surface modes
in which the damping associated to individual particles is
suppressed, can have profound implications in future nanoplasmonic
research~\cite{Carron86, ZouJanel04}. These predictions have been
confirmed in 1D plasmonic crystals~\cite{Kall05} and, very recently,
in 2D crystals~\cite{Grigorenko08, Barnes08, Crozier08} by the
observation of sharp resonances in dark-field scattering and
extinction spectra. Also the coupling between nanoparticles in 2D
crystals through guided optical modes has been recently
reported~\cite{Linden01, Felidj02}. However, the important question
of how light emitters can couple to lattice surface modes and how
their emission is influenced by this coupling still remains
unexplored.

In this article, we demonstrate a strong modification and
directional enhancement of the fluorescence of emitters in the
proximity of plasmonic crystals of nanoantennas. This modified
emission is the result of the electrodynamical coupling of the
emitters to surfaces modes that result from the interaction of LSPRs
with photonic states. Moreover, we prove that the condition for the
existence of these mixed modes, namely the embedding of the
nanoparticle array in a fully symmetric dielectric environment, is
less stringent than previously believed~\cite{Barnes08}. We support
our conclusions with Finite-Different-Time-Domain (FDTD)
calculations.

\begin{figure}
\centerline{\scalebox{0.6}{\includegraphics{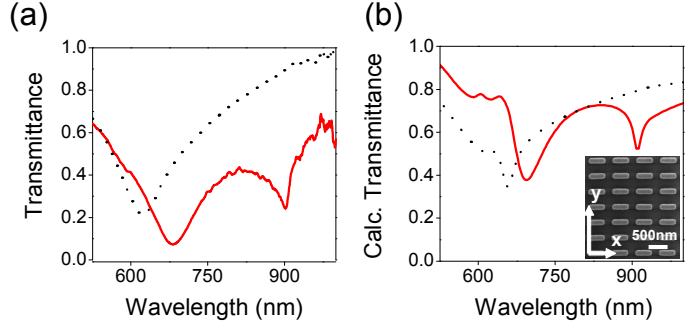}}}
\caption{(a) Measurements and (b) FDTD calculations of the
transmittance through a 2D plasmonic crystal of gold nanoantennas on
top of a glass substrate (black-dotted lines) and through a similar
plasmonic crystal covered by a 50 nm thick PVB layer (red-solid
lines). Inset in (b): SEM image of a plasmonic
crystal.}\label{Trans}
\end{figure}

Large arrays (3 $\times$ 3 mm$^{2}$) of gold nanoantennas were
fabricated by nanoimprint onto a glass substrate with refractive
index 1.52 at $\lambda = 700$ nm. A scanning electron microscope SEM
image of the top of the sample is shown in the inset of
Fig.~\ref{Trans}(b). The height of the nanoantennas is $38 \pm 2$
nm. The grating constants are $a_{x} = 600 \pm 15$ nm and $a_{y} =
300 \pm 15$ nm. The nanoantennas are rectangular nanorods of rounded
corners, with long (short) axis of $450\pm 10$ nm ($130 \pm 10$ nm)
along the x (y) direction. An active layer of thickness $50 \pm 10$
nm was spun onto the array. This layer consists of fluorescent
molecules (ATTO 680) dispersed into a polyvinyl butyral (PVB) matrix
with a concentration of $10^{-5}$ M. The top layer has a refractive
index of 1.48 at $\lambda = 700$ nm.

Arrays of metallic nanoparticles may interact through coherent
scattering processes. Diffractive coupling between LSPRs of the
nanoparticles and photonic modes of the array occurs whenever the
wavelength of the scattered light by LSPRs approaches the
interparticle distance. The complex polarizability of an isolated
particle, $\alpha$, no longer describes such system. The array of
nanoparticles, in the framework of the coupled dipole approximation,
can be modelled by an effective polarizability $\alpha_{\rm
eff}=\alpha/(1-\alpha S)$, where and $S$ is a term representing the
retarded dipole sum resulting from coherent scattering in the
nanoparticle array~\cite{ZouJanel04}. Scattering resonances, which
give rise to lattice surface modes, arise whenever the real
components of $\alpha^{\rm-1}$ and $S$ are equal~\cite{ZouJanel04,
Abajo06}. A reduction of the single particle damping also occurs
because of the relative compensation of imaginary components of
$\alpha^{\rm-1}$ and $S$. This reduction of the damping produces
sharp resonances~\cite{Kall05, Grigorenko08, Barnes08, Crozier08}.
As we will show later, light emitters can couple very efficiently to
these modes modifying drastically their emission characteristics and
exhibiting narrow bands of enhanced emission.

A lattice resonance is displayed in Fig.~\ref{Trans}(a) near to
$\lambda=900$ nm, where the red-solid line represents a measurement
of the zero-order transmission  at normal incidence through the
plasmonic crystal of nanoantennas on glass covered by the PVB layer.
The measurement is normalized by the transmission through a bare
substrate covered by a similar PVB layer. We illuminated the sample
with a collimated beam from an halogen lamp with the polarization
parallel to the short axis of the nanoantennas. The broad reduction
of the transmittance around 680 nm corresponds to an enhanced
extinction resulting from the excitation of the half-wave LSPR in
each individual nanoantenna. The asymmetric and narrower resonance
around 900 nm arises from the diffractive coupling of the
nanoantennas arranged in the plasmonic crystal. It has been
highlighted the importance of surrounding the nanoparticles by a
homogeneous dielectric environment in order to obtain an effective
diffractive coupling~\cite{Barnes08}. Fig.~\ref{Trans}(a) shows the
transmittance measured through a similar array without top PVB layer
(black-dotted line). The single particle LSPR is shifted to 620 nm
due to the reduced permittivity of the surrounding medium. The most
pronounced difference compared to the measurement of the sample with
PVB layer is the disappearance of the lattice surface mode, as a
result of the inefficient diffractive coupling in the inhomogeneous
environment. We stress that the nanoantenna array covered by PVB is
not immersed in a fully symmetric dielectric environment. Instead,
there is a small difference in refractive index between the glass
substrate and the polymer layer, and this layer has a finite
thickness of only 50 nm. Therefore, our measurements demonstrate
that the condition for the existence of these modes, namely, an
homogeneous dielectric environment around the nanoparticle array, is
less stringent than believed.

\begin{figure}
\centerline{\scalebox{0.5}{\includegraphics{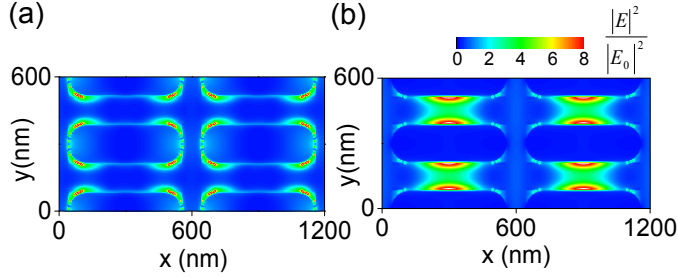}}}
\caption{Near field intensity enhancement in a 2D plasmonic crystal
of nanoantennas on a glass substrate covered by a PVB layer. The
field is calculated on the plane intersecting the nanoantennas at
their middle height (z = 20 nm). The calculations are at (a)
$\lambda=695$ nm, and (b) $\lambda=905$ nm, which correspond to the
LSPR and to the lattice surface mode respectively. }\label{FDTD}
\end{figure}

The main features in the measurements of Fig.~\ref{Trans}(a), namely
the LSPR and the lattice resonance, are well reproduced by FDTD
calculations shown in Figure~\ref{Trans}(b). The shoulder around
$\lambda=600$ nm related to intraband transitions in gold is much
less pronounced, although still visible, in the measurements.
Calculations of the electric near field enhancement, i.e., the near
field intensity normalized by the incident intensity, are shown in
Fig.~\ref{FDTD} for two different wavelengths, (a) 695 nm and (b)
905 nm. At $\lambda=695$ nm, i.e., the wavelength of the LSPR, the
field is only enhanced at the edges of nanoantennas. On the
contrary, the diffractive coupling of nanoantennas leading to the
lattice surface mode at $\lambda=905$ nm produces a drastic
redistribution of the near field intensity with a much larger field
enhancement in between the nanoantennas. This lattice surface mode
represents a new decay channel for dye molecules in the proximity of
the nanoantenna crystal.

\begin{figure}
\centerline{\scalebox{0.6}{\includegraphics{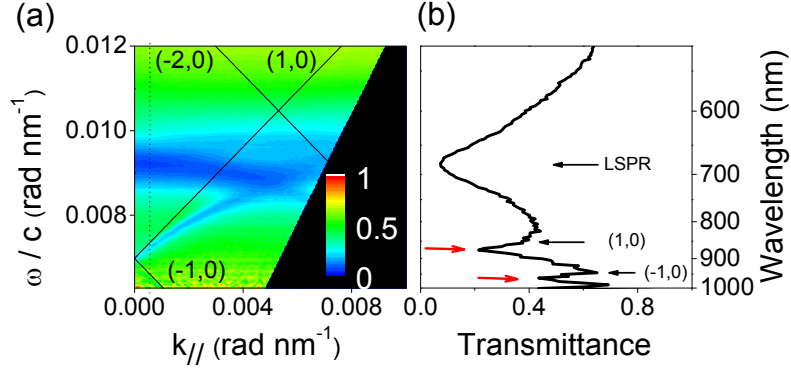}}}
\caption{(a) Transmittance of light polarized parallel to the short
axis of the nanoantennas as a function of the normalized frequency,
$\omega/c$, and the wave number ${k_\|}$. The solid lines represent
the Rayleigh anomalies. (b) Transmittance spectrum at
${k_\|}=5.47\times 10^{-4}$ rad~nm$^{-1}$, i.e., the wave vector
indicated by the dotted vertical line in (a). Black arrows indicate
the Rayleigh condition and the LSPR, while the red arrows indicate
the lattice surface resonances.}\label{Disprel}
\end{figure}

To obtain the dispersion relation of these modes, we have measured
the zero-order transmission spectra as a function of the incident
angle $\theta_{\rm in}$ in the range $0^\circ-50^\circ$. These
measurements are displayed in Fig.~\ref{Disprel}(a) as a function of
${k_\|}$, where ${\textbf k_\|}={k_\|}{\hat{x}}={k_{\rm
0}}\sin(\theta_{\rm in}){\hat{x}}$, and ${k_{\rm 0}}= {\omega}/{c}$.
The so-called Rayleigh anomalies, or the condition at which a
diffracted order becomes grazing to the grating surface, have been
also plotted in Fig.~\ref{Disprel} with black lines. These anomalies
are given by the equation $k_{\rm out}^2=k_{\rm in}^2 \sin^2
\theta_{\rm in} + m_1^2 ({2\pi}/{a_x})^2+m_2^2 ({2\pi}/{a_y})^2 +
2k_{\rm in} \sin \theta_{\rm in} m_1 ({2\pi}/{a_x})$, where $k_{\rm
in}$ and $k_{\rm out}$ are the incident and scattered wave numbers
associated to the parallel component of the wave vector to the plane
of the array, and $(m_1,m_2)$ are integers defining the diffraction
order. The Rayleigh anomalies displayed in Fig.~\ref{Disprel} were
calculated taking $m_2=0$, $a_x=600$ nm and a refractive index
$n=1.50$, which corresponds to the average between the glass and PVB
indices. The LSPR at $\omega/c=0.0092$ rad~nm$^{-1}$ is manifest in
Figs.~\ref{Disprel}(a) and (b) by the broad reduction in
transmittance. The localized nature of this resonance, as it was
apparent in the calculation of Fig.~\ref{FDTD} (a), is also evident
from its flat dispersion relation. The lattice surface modes
resulting from the diffractive coupling of LSPRs appear as narrow
bands of reduced transmittance at lower frequencies than the
Rayleigh anomaly. Let us focus on the mode associated to the (1,0)
diffraction order. As the frequency increases towards the LSPR
frequency, the wave number of the surface mode deviates from the
Rayleigh anomaly and also the width of the resonance
increases~\cite{Barnes08}. The dispersive behavior of these modes is
similar to that of surface plasmon polaritons propagating on
metallic gratings.

In the following, we focus on the emission of the dye molecules
embedded in the PVB layer and their coupling to the lattice surface
modes in the plasmonic crystal. The one photon fluorescence
enhancement of an emitter excited below saturation is defined as the
emitted intensity ($I$) normalized by the intensity of the emitter
in the absence of the plasmonic crystal ($I_{0}$). For dipoles
randomly oriented in space, this enhancement is given by
\begin{equation}
\frac{I}{I_{0}}=\frac{\eta({\textbf r}_0,\omega_{em})}{\eta_0
({\textbf r}_0,\omega_{em})} \frac{\left|\textbf E({\textbf
r}_0,\omega_{abs})\right|^2}{\left|\textbf {E}_{0}({\textbf
r}_0,\omega_{abs})\right|^2} \;, \label{enhancement}
\end{equation}
where ${\textbf E}$ and ${\textbf E}_{0}$ are the local and incident
electric fields at the pump frequency $\omega_{abs}$ and $\eta$ and
$\eta_0$ are the quantum yield of molecules with and without the
presence of the nanoantennas. The term ${\left|\textbf
E\right|^2}/{\left|\textbf {E}_{0}\right|^2}$ in
Eq.~(\ref{enhancement}) represents the pump enhancement, or the
increase of the fluorescence of an emitter located at ${\textbf
r}_0$ due to the local enhancement of the electromagnetic field at
the frequency of excitation. The term ${\eta}/{\eta_0}$ in
Eq.~(\ref{enhancement}) corresponds to the increase of the emission
due the enhancement of the local density of optical states (LDOS) at
${\textbf r}_0$ to which the emitter can decay emitting radiation of
frequency $\omega_{em}$.

\begin{figure}
\centerline{\scalebox{0.6}{\includegraphics{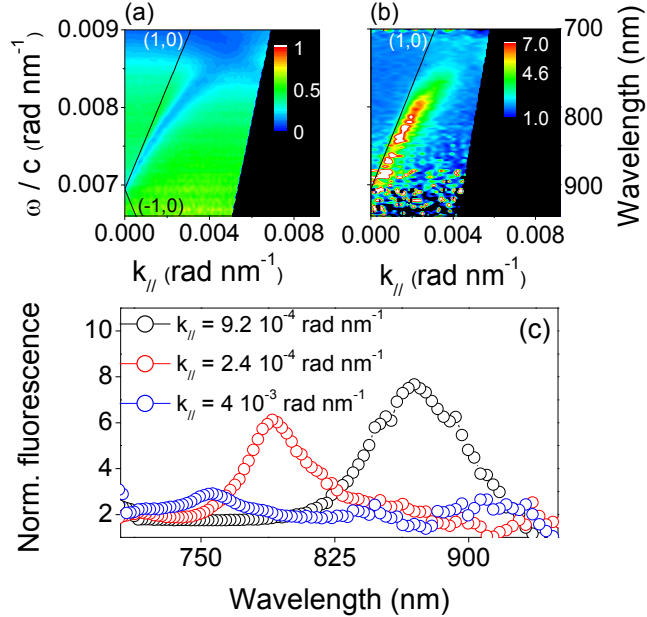}}}
\caption{(a) Transmittance as in Fig.~\ref{Disprel}(a). (b)
Fluorescence enhancement or the fluorescence of dye molecules in the
proximity of a nanoantenna plasmonic crystal normalized by the
fluorescence of dye molecules on an unpatterned substrate. (c)
Fluorescence enhancement spectra at three different values of
${k_\|}$.}\label{PL}
\end{figure}

To investigate the emission of sources coupled to plasmonic
crystals, we have performed fluorescence measurements of ATTO 680
molecules embedded in the PVB layer. ATTO 680 is a photostable dye
with a maximum absorption cross section at $\lambda=$ 680 nm and a
fluorescence spectrum peaked at 700 nm and extending until 950 nm.
The dye molecules were excited at a fixed angle of illumination
($\theta_{\rm in} = 50^\circ$) using a laser diode emitting at
$\lambda=685$ nm. At this wavelength ATTO 680 absorbs very
efficiently and the unfocused pump, of beam size $\sim 1$ mm,
excites uniformly the molecules over a large area. The laser power,
set to 1 mW, was weak enough to neglect saturation effects. We
selected the fluorescence emission with polarization parallel to the
short axis of the nanoantennas. The angular detection range of the
fluorescence was from $0^\circ$ to $40^\circ$ in the plane of
incidence. The results of the fluorescence measurements are
summarized in Figure~\ref{PL}. Figure~\ref{PL}(b) represents the
fluorescence intensity measured on the array of nanoantennas
normalized to the fluorescence of a similar layer of PVB and dye
molecules on top of an unpatterned substrate. The normalized
fluorescence is represented as a function of ${k_\|}$ and of the
normalized frequency $\omega/c$. For a direct comparison with the
variable-angle transmittance measurements, we display these
measurements in Fig.~\ref{PL}(a) in the same range of wave numbers
and frequencies than the fluorescence measurements. In
Fig.~\ref{PL}(c) we display three normalized fluorescence spectra
obtained from Fig.~\ref{PL}(b) at different values of ${k_\|}$. It
is evident from Figs~\ref{PL}(a) and (b) that the highest
enhancement of the fluorescence (up to seven-fold enhancement) takes
place on a narrow band that reproduces the dispersive behavior of
the lattice surface mode resulting from the diffractive coupling of
LSPRs.

\begin{figure}
\centerline{\scalebox{0.6}{\includegraphics{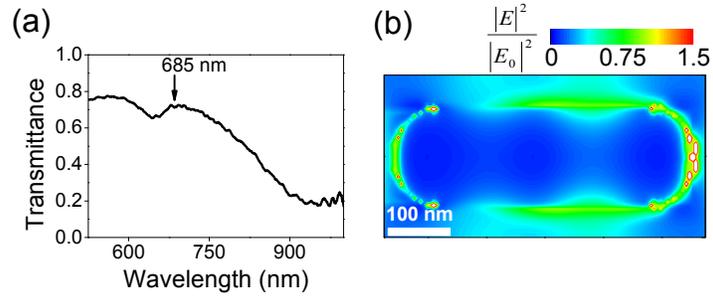}}}
\caption{(a) Transmittance spectrum through a plasmonic crystal of
nanoantennas for incident angle $\theta_{\rm in}=50^\circ$ and
polarization parallel to the long axis of nanoantennas. (b) FDTD
calculation of the near field intensity enhancement around one
nanoantenna at a wavelength of 685 nm and for an angle of incidence
$\theta_{\rm in}=50^\circ$.}\label{Pump}
\end{figure}

To rule out a possible contribution of a local pump enhancement to
the fluorescence emission, we have measured the transmittance
spectrum (Fig.~\ref{Pump}(a)) through the plasmonic crystal for an
incident angle $\theta_{\rm in} = 50^\circ$, i.e., the angle of
incidence of the pump laser. The polarization of the light was set
parallel to the long axis of the nanorods, as was for the excitation
beam during the fluorescence measurements. The reductions of the
transmittance around 900 nm and 650 nm in Fig.~\ref{Pump}(a) are
attributed to the excitation of the second and third order LSPRs
along the long axis of the nanoantennas. The most relevant feature
in Fig.~\ref{Pump}(a) for our measurements is the high transmittance
at $\lambda=685$ nm, which indicates the absence of resonant
scattering by the nanoantennas at the wavelength of the pumping
laser. This absence of resonant scattering leads to a negligible
local field enhancement as can be appreciated in the FDTD
calculation displayed in Fig.~\ref{Pump}(b).

Being discarded a mayor contribution of a local field enhancement of
the pump to the fluorescence, the large directional enhancement of
the emission in Fig~\ref{PL} is attributed to the first factor of
Eq.~(\ref{enhancement}), i.e., to an increase of the LDOS associated
to lattice surface modes to which the excited dye molecules can
decay radiatively. The delocalized nature of these modes, with a
large field intensity distributed in between the nanoantennas, is
advantageous over the LSPRs for an efficient fluorescence emission
because of the following reasons: (i) a larger number of dye
molecules can efficiently couple to this mode due to its large
extension; (ii) The average distance between fluorescent molecules
that couple to the mode and the metal is larger, which reduces the
metal induced fluorescence quenching, and (iii) the dispersion
relation of surface mode is close to the Rayleigh anomaly (black
line in Fig.~\ref{PL}(b)), which means that a small extra momentum
is required to efficiently couple out the fluorescence to radiative
modes. This last point is clearly visible in Figs.~\ref{PL}(b) and
(c), where the fluorescence enhancement at the lattice resonance
gradually decreases for increasing ${k_\|}$ or as the resonance
deviates from the Rayleigh anomaly. It is important to notice that
only a weak enhancement of the emission is observed at the LSPR,
i.e., the enhancement is a factor of two around at $\lambda$=700 nm,
which highlights the higher efficiency of lattice surface modes over
LSPRs for light extraction in the proximity of plasmonic crystals.

In summary, we have demonstrated experimentally that lattice surface
modes resulting from the coupling of localized surface plasmon
resonances in plasmonic crystals of nanoantennas are responsible of
a strong modification of the emission properties of dye molecules in
proximity of the crystal. Enhanced fluorescence, due to the
increased local density of optical states to which dye molecules can
decay, has been measured within narrow spectral and angular bands.
This observation introduces new perspectives in the field of surface
enhanced fluorescence for, e.g., microscopy imaging, biology, and
optical devices.

We are thankful to M. Verschuuren, Y. Zhang and K. Catchpole for
assistance during the fabrication of the samples and with the
calculations, and to F.J. Garc\'ia de Abajo for stimulating
discussions. This work was supported by the Netherlands Foundation
Fundamenteel Onderzoek der Materie (FOM) and the Nederlandse
Organisatie voor Wetenschappelijk Onderzoek (NWO), and is part of an
industrial partnership program between Philips and FOM.

\bibliographystyle{unsrt}
\bibliography{bibliography}

\begin{thebibliography}{100}

\bibitem{Maier04} S.A. Maier and H.A. Atwater,  J. Appl. Phys. \textbf{98}, 011101 (2005), and references therein.
\bibitem{Lal07} S. Lal, S. Link, and N.J. Halas, Nature Photonics \textbf{1}, 641 (2007), and references therein.
\bibitem{Muhlschlegel05} P. M\"{u}hlschlegel, H.-J. Eisler, O.J.F. Martin, B. Hecht, and D.W. Pohl, Science \textbf{308}, 1607 (2005).
\bibitem{Lakowicz05} J.R. Lakowicz, Anal. Biochem. \textbf{337}, 171 (2005).
\bibitem{Sandoghdar06} S. K\"{u}hn, U. H\r{a}kanson, L. Rogobete, and V. Sandoghdar, Phys. Rev. Lett. \textbf{97}, 017402 (2006).
\bibitem{Mertens06} H. Mertens, J.S. Biteen, H.A. Atwater, and A. Polman, Nano Lett. \textbf{6}, 2622 (2006).
\bibitem{Krenn07} S. Gerber, F. Reil, U. Hohenester, T. Schlagenhaufen, J.R. Krenn, and A. Leitner, Phys. Rev. B \textbf{75}, 073404 (2007).
\bibitem{Tam07} F. Tam, G.P. Goodrich, B.R. Johnson, and N.J. Halas, Nano Lett. \textbf{7}, 496 (2007).
\bibitem{Muskens07} O. Muskens, V. Giannini, J.A. S\'{a}nchez-Gil, and J. G\'{o}mez Rivas, Nano Lett. \textbf{7}, 2871 (2007).
\bibitem{Bakker08} R.M. Bakker, H.K. Yuan, Z.T. Liu, V.P. Drachev, A.V. Kildishev, V.M. Shalaev, R.H. Pedersen, S. Gresillon, and A. Boltasseva, Appl. Phys. Lett. \textbf{92}, 043101 (2008).
\bibitem{Ford84} G.W. Ford and W.H. Weber, Phys. Rep. \textbf{113}, 195 (1984).
\bibitem{Feldmann05} E. Dulkeith, M. Ringler, T.A. Klar, J. Feldmann, A. Mu\~{n}oz Javier, and W.J. Parak, Nano Lett. \textbf{5}, 585 (2005).
\bibitem{Novotny06} P. Anger, P. Bharadwaj, and L. Novotny, Phys. Rev. Lett. \textbf{96}, 113002 (2006).
\bibitem{Polman08} A. Polman, Science \textbf{322}, 868 (2008).
\bibitem{Soref07} J.B. Khurgin, G. Sun, and R.A. Soref, J. Opt. Soc. Am. B \textbf{24}, 1968 (2007).
\bibitem{Carron86} K.T. Carron, W. Fluhr, M. Meier, A. Wokaun, and H.W. Lehmann, J.Opt. Soc. Am. B \textbf{3}, 430 (1986).
\bibitem{ZouJanel04} S. Zou, N. Janel, and G.C. Schatz, J. Chem. Phys. \textbf{120}, 10871 (2004).
\bibitem{Kall05} E.M. Hicks, S. Zou, G.C. Schatz, K.G. Spears, R.P. van Duyne, L. Gunnarsson, T. Rindzevicius, B. Kasemo, and M. K\"{a}ll, Nano Lett. \textbf{5}, 1065 (2005).
\bibitem{Grigorenko08} V.G. Kravets, F. Schedin, and A.N. Grigorenko, Phys. Rev. Lett. \textbf{101}, 087403 (2008).
\bibitem{Barnes08} B. Augui\'{e} and W.L. Barnes, Phys. Rev. Lett. \textbf{101}, 143902 (2008).
\bibitem{Crozier08} Y. Chu, E. Schonbrun, T. Yang, and K.B. Crozier, Appl. Phys. Lett. \textbf{93}, 181108 (2008).
\bibitem{Linden01} S. Linden, J. Kuhl, and H. Giessen, Phys. Rev. Lett. \textbf{86}, 4688 (2001).
\bibitem{Felidj02} N. F\'{e}lidj, J. Aubard, G. L\'{e}vi, J.R. Krenn, G. Schider, A. Leitner, and F.R. Aussenegg, Phys. Rev. B \textbf{66}, 245407 (2002).
\bibitem{Abajo06}F. J. Garc\'{i}a de Abajo, J. J. S\'{a}enz, I. Campillo, and J. S. Dolado, Opt. Express \textbf{14}, 7 (2006).

\end{thebibliography}


\newpage

\newpage

\newpage

\newpage

\newpage

\end{document}